\documentclass[final,authoryear,3p,times,twocolumn]{elsarticle}
\biboptions{longnamesfirst,angle,semicolon}
\usepackage{amssymb}
\usepackage{txfonts}
\usepackage{natbib}
\usepackage{graphicx,epsfig,fancyhdr,rotating,amsmath,natbib}
\usepackage{aas_macros}

\journal{New Astronomy}

\begin{document}

\begin{frontmatter}

\title{Interstellar polarization and extinction toward the Recurrent Nova T CrB\tnoteref{1}}
\tnotetext[1]{Based on data collected with 2-m RCC telescope at Rozhen National Astronomical Observatory.}

\author[]{Yanko Nikolov\corref{1}}
\ead{ynikolov@nao-rozhen.org}\cortext[1]{Corresponding author.}

\address{
    Institute of Astronomy and National Astronomical Observatory, Bulgarian Academy of Sciences\unskip, Tsarigradsko Shose 72\unskip, Sofia\unskip, BG-1784\unskip, Bulgaria}

\begin{abstract}
	Spectropolarimetry is a powerful tool for diagnostic of interstellar matter and gives information on the geometry of the ejected material after the novae outbursts. In this paper are presented spectropolarimetric observations of the recurrent nova T CrB at quiescence obtained with FoReRo2 attached to the Cassegrain focus of the 2.0m RCC telescope of the Bulgarian Rozhen National Astronomical Observatory. The interstellar polarization toward T CrB was estimated using the field stars method. The spectropolarimetric observations were obtained from February 2018 to August 2021. In the wavelength range from 4800~\AA~ to 8200~\AA~the maximum of the degree of linear polarization is $P_{max}(obs)(\%) = 0.46 \pm 0.01$ at $\lambda \approx 5200$ \AA. The position angle is $P.A._{obs}=100^{\circ}.8 \pm 0^{\circ}.9$.  During the observations, there is no intrinsic polarization in T CrB, and the derived values represent interstellar polarization. The polarization toward T CrB is due to the foreground interstellar dust located at the distance up to $\approx$ 400 pc. Based on the degree of polarization the interstellar extinction toward the T CrB is $E_{B-V} \approx 0.07$.

\end{abstract} 




\end{frontmatter}



\section{Introduction}

T CrB (HD 143454) is a symbiotic recurrent nova system. 
The Recurrent Novae (RNe) are a heterogeneous group of binary systems that have more than one recorded nova outburst. Two nova-like outbursts of T CrB in 1868 and 1946 are observed.
There are currently 10 confirmed RNe in the Galaxy \citep{2010ApJS..187..275S}. The presence of a massive, accreting white dwarf in the RNe systems makes them potential progenitors of Type Ia supernovae. T CrB consist of a massive white dwarf with $M_{WD}$ = 1.2-1.37 $M_{\odot}$ (\citet{1998MNRAS.296...77B}, \citet{2004A&A...415..609S}) and red giant with $M_{RG} \approx 1.12 M_{\odot}$ \citep{2004A&A...415..609S}. The orbital period of the binary is $227^{d}.57$ \citep{2000AJ....119.1375F}.\\ 
Recurrent Novae are classified into long and short period systems \citep{2013A&A...559A.121A}. The donor of the long period systems is a red giant: RS Oph, T CrB, V3890 Sgr and V745 Sco. 
The short period systems are further divided into U Sco and T Pyx groups based on the outburst and quiescent properties \citep{2013A&A...559A.121A}. Reviews on the properties of RNe can be found in \citet{2008ASPC..401...31A} and \citet{2010ApJS..187..275S}.\\
Similar to RS Oph, T CrB is a symbiotic recurrent nova system. An intrinsic degree of polarization was observed two days after the last outburst of the recurrent nova RS Oph \citep{2021ATel14863....1N}. The intrinsic polarization is $\approx$ 1\% with well visible strong depolarization effects in $H_{\alpha}$ and He I 5876 emission lines. The position angle is aligned with the highly collimated outflows detected during the past outburst in 2006 \citep{2008ApJ...688..559R}. To obtain an intrinsic degree of polarization and position angle of the recurrent nova T CrB shorter after the upcoming outburst \citep{2020ApJ...902L..14L}, it is necessary to obtain the interstellar polarization toward T CrB.\\
In this study are presented spectropolarimetric observations of T CrB, obtained with the 2-Channel-Focal-Reducer Rozhen (FoReRo2) attached to the 2.0m RCC telescope of the Bulgarian Rozhen National Astronomical Observatory.

\section{Observations and Data Analysis}

Spectropolarimetric observations of T CrB were secured with the 2-Channel-Focal-Reducer Rozhen (FoReRo2) similar to that described by \citet{2000KFNTS...3...13J}, 
attached to the Cassegrain focus of the 2.0m RCC telescope 
of the Bulgarian Rozhen National Astronomical Observatory. 
A Super-Achromatic (in the range 
3800 \AA -- 7900 \AA) True Zero-Order Waveplate~5 retarder (APSAW-5)\footnote{http://astropribor.com/waveplates/} has been added to FoReRo2. 
Polarized spectra are obtained at 8 retarder angles: $0^{\circ}$, $22.5^{\circ}$, $45^{\circ}$, $67.5^{\circ}$, $90^{\circ}$, $112.5^{\circ}$, $135^{\circ}$ and $157.5^{\circ}$. 
A beam swapping technique is used (\citet{2009PASP..121..993B}, \citet{2019AcA....69..361N}) to minimize instrumental polarization.
Instrumental polarization is corrected using a standard star with zero degree of polarization.
The offset between position angle in the celestial and instrumental polarization is corrected using 
strongly polarized standard stars. 
The observations were made from 2018 February 17 to 2021 August 08, covering the orbital period of T CrB. 
Polarized spectra of T CrB, the stars from the field of T CrB, and standard stars 
with high (HD 204827 and HD 161056, \citet{1992AJ....104.1563S}) and zero degrees of polarization (HD 212311 and HD 154892, \citet{1990AJ.....99.1243T}) were obtained. Polarization standards and the stars from the field of T CrB were observed with the same instrumental setup as T CrB on the same nights. The spectra were reduced using IRAF \citep{1993ASPC...52..173T} in the standard way including bias removal and wavelength calibration. 
The journal of observations is given in Table \ref{tab.obsjournal}. The table contains: objects, the date (in format YYYYMMDD), UT of the middle of the observation, the total exposure time for all retarder angles, observed high and zero polarization standards, angular distance from T CrB and distance. The last column represents 
observed degree of polarization and position angle in synthetic V filter.

\begin{table*}
\centering
\caption{Journal of observations of T CrB and the stars from the field of T CrB.}
\begin{tabular}{ l  c  c  c  c c c c c c }
\hline
\hline
Object & Date   & UT & $Exp.^{a}$(s)  & Standards & Angular   & $Distance^{b}$ & $P_{V}$ & $P.A._{V}$ \\
       &        &    &                &           & distance  & pc       &  (\%)   & [deg.]     \\
\hline
		&2018-02-17~&  ~01:24&   960   & 1   &    &			   & $0.35 \pm 0.03^{c}$     &$99.5 \pm 2.0^{c}$     \\
		&2018-10-07~&  ~17:41&   720   & 2   &    &			   & $0.44 \pm 0.04^{c}$     &$101.6 \pm 2.6^{c}$    \\
		&2019-06-30~&  ~22:11&   1920  & 1   &    &			   & $0.43 \pm 0.02^{c}$     &$101.1 \pm 1.8^{c}$    \\
		&2019-07-02~&  ~21:09&   1440  & 1   &    &			   & $0.36 \pm 0.02^{c}$     &$102.0 \pm 2.0^{c}$    \\
		&2019-10-25~&  ~17:21&   960   & 2   &    &			   & $0.41 \pm 0.02^{c}$     &$103.7 \pm 1.8^{c}$    \\
\bf T CrB 	&2020-02-02~&  ~01:27&   240   & 1   & -- & $ 887^{+18}_{-29} $    & $0.42 \pm 0.04^{c}$     &$99.5 \pm 2.9^{c}$     \\
		&2020-04-18~&  ~02:11&  144    & 1   &    &			   & $0.49 \pm 0.10^{c}$     &	$104.6 \pm 5.4^{c}$  \\
		&2020-04-19~&  ~20:22&   240   & 1   &    &			   & $0.48 \pm 0.03^{c}$     &	$105.2 \pm 1.8^{c}$  \\
		&2020-04-28~&  ~19:46&  192    & 1   &    &			   & $0.41 \pm 0.07^{c}$     &$95.3 \pm 5.8^{c}$      \\
		&2021-05-04~&  ~20:41&  1440   & 1   &    &			   & $0.42 \pm 0.02^{c}$     &$97.2 \pm 1.4^{c}$     \\
		&2021-08-08~&  ~19:58&  960    & 1   &    &			   & $0.41 \pm 0.03^{c}$     &$98.9 \pm 2.4^{c}$     \\
\hline
              & 2020-02-02~ &  ~01:38&   480 &  1 &	    &			   & $0.41 \pm 0.03^{c}$    & $104.0 \pm 2.2^{c}$  \\
\bf HD 143256 & 2020-04-19~ &  ~20:31&   960 &  1 &  14'    &$ 613^{+5}_{-4} $     & $0.40 \pm 0.03^{c}$    & $106.5 \pm 1.9^{c}$	    \\
	      & 2020-04-28~ &  ~20:12&   480 &  1 &	    &			   & $0.38 \pm 0.03^{c}$    &  $96.1 \pm 2.8^{c}$     \\
\hline
	        & 2020-02-02~ &  ~01:53&   240 & 1 &	    &			   & $0.33 \pm 0.07^{c}$  & $104.0 \pm 4.7^{c}$      \\
\bf BD +26 2759 & 2020-04-19~ &  ~20:54&   320 & 1 &  22.3' &$ 1066^{+26}_{-26} $  & $0.48 \pm 0.06^{c}$  &  $107.4 \pm 3.3^{c}$     \\
	        & 2020-04-28~ &  ~19:57&   360 & 1 &	    &			   & $0.43 \pm 0.06^{c}$  &  $95.5 \pm 3.5^{c}$      \\
\hline
\bf HD 143161 & 2020-02-02~ &  ~02:10&   480 & 1 & 23.6' &$ 759^{+9}_{-9} $	   &$0.40 \pm 0.02^{c}$  & $103.1 \pm 1.8^{c}$   \\
              & 2020-04-19~ &  ~21:08&   400 & 1 &	 &			   &$0.38 \pm 0.04^{c}$	 & $107.3 \pm 2.4^{c}$       \\
\hline
\bf  2MASS               & 2020-02-02~ &  ~02:32&   1200 & 1 & 	         & 	              &	$0.36 \pm 0.06^{c}$  &  $101.5 \pm 5.9^{c}$      \\
\bf  16005644            & 2020-04-19~ &  ~21:36&   1200 & 1 & 20' 	 & $ 727^{+8}_{-7} $  &	$0.40 \pm 0.07^{c}$  &  $75.5 \pm 6.4^{c}$      \\
\bf  +2550020            &&&&&&&& \\
\hline
\bf 2MASS                          &&&&&&&& \\
\bf 16000301               & 2020-02-02~ &  ~02:58&   1200 & 1 & 7.4'      &$ 919^{+13}_{-11} $ & $0.32 \pm 0.10^{c}$  &$95.9 \pm 26.0^{c}$      \\
\bf +2555003         &&&&&&&& \\ 
\hline
\bf HD 142762 & 2021-08-08~ &  ~20:26&   960 &  1 &  $0.94^{\circ}$	&$ 278^{+2}_{-2} $ & $0.38 \pm 0.01^{c}$   & $99.4 \pm 1.3^{c}$      \\	      
\hline
\bf HD 142053 & 2021-08-08~ &  ~20:57&   480 &  1 &  $1.91^{\circ}$	&$ 217^{+1}_{-1} $ & $0.47 \pm 0.01^{c}$  &$110.6 \pm 1.0^{c}$     \\ 
\hline
\bf HD 138749 & 2021-08-08~ &  ~21:19&   72 &  1 &   $7.96^{\circ}$     &$ 121^{+2}_{-3} $ & $0.30 \pm 0.01^{c}$  &$93.7 \pm 0.8^{c}$     \\

\hline

\label{tab.obsjournal}
\end{tabular} 
\\
Note: $^{a}$ - Total exposure time; 
(1) - high polarization standard HD 161056 \& zero polarization standard HD 154892; 
(2) - high polarization standard HD 204827 \& zero polarization standard - HD 212311;
 $^{b}$ Distance based on Gaia DR3 \citep{2021AJ....161..147B}. 
 
\end{table*}

\section{ Results }
\label{results}

\subsection{Observed degree of polarization and position angle of T CrB}

\begin{figure}[htb]
    \begin{center}
       \includegraphics[width=0.5\textwidth, angle=0]{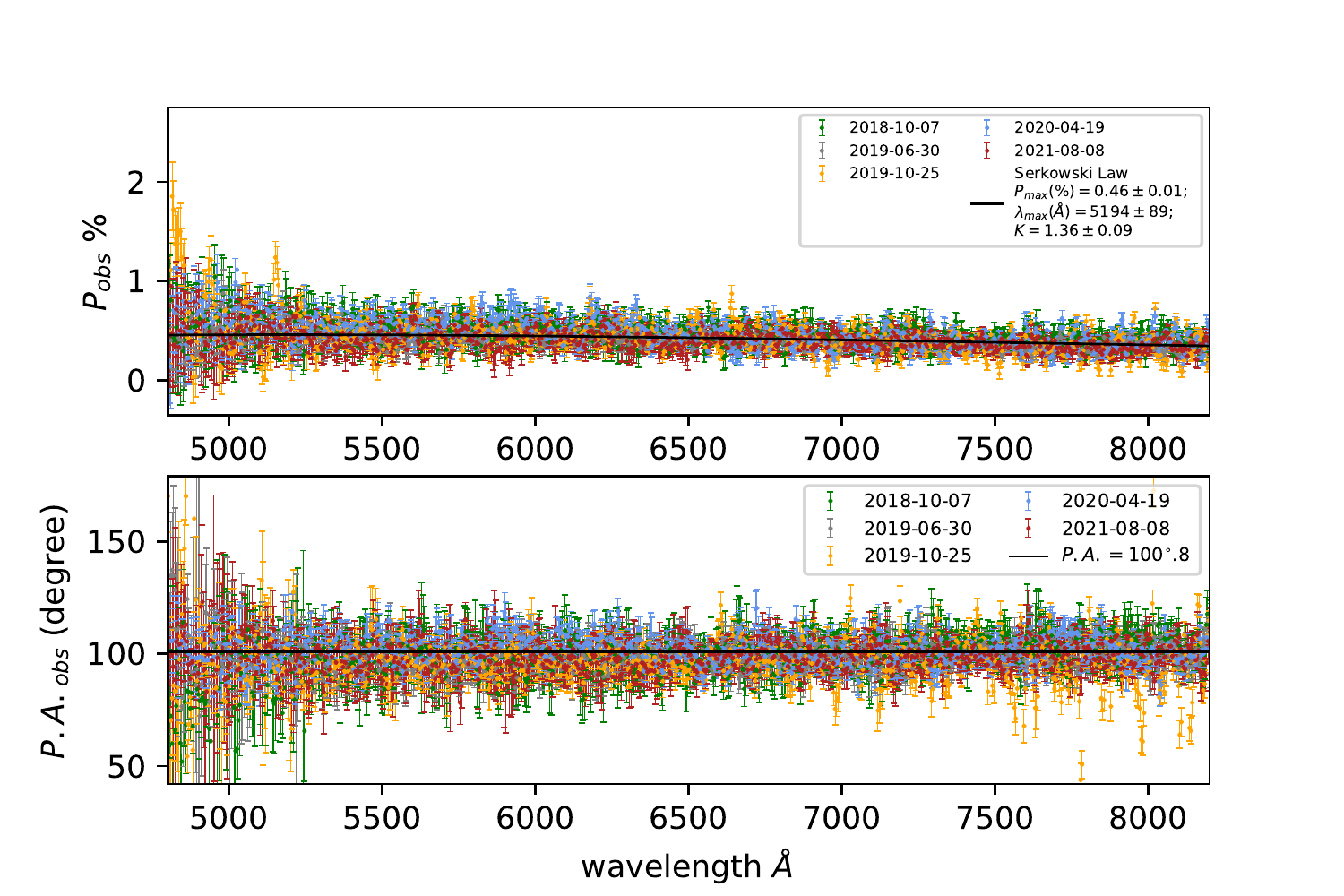}
     \end{center}
         \caption[]{Degree of polarization and Position Angle of T CrB. Black curve represents fit with Serkowski's law with coefficients $P_{max}$=0.46, K=1.36 and $\lambda_{max}$=5194\AA.}
	 
\label{pol.obs.tcrb}	 
\end{figure}

The observed degree of polarization and position angle in the wavelength range from 4800~\AA~to 8200~\AA~are plotted on Fig.\ref{pol.obs.tcrb}. The maximum of the degree of polarization is $P_{max}(obs)(\%) = 0.46 \pm 0.01$ at $\lambda \approx 5200$ \AA.
The angle of polarization has a flat behavior with no visible wavelength dependence and it has a value of $P.A._{obs} = 100^{\circ}.8 \pm 0^{\circ}.9$, where $P.A._{obs}$ represents an average value of all observed position angles of T CrB in synthetic V filter. The degree of polarization can be fit with Serkowski's law with coefficient $P_{max}$=0.46, K=1.36 and $\lambda_{max}$=5194\AA.

\begin{figure}[htb]
    \begin{center}
       \includegraphics[width=0.50\textwidth, angle=0]{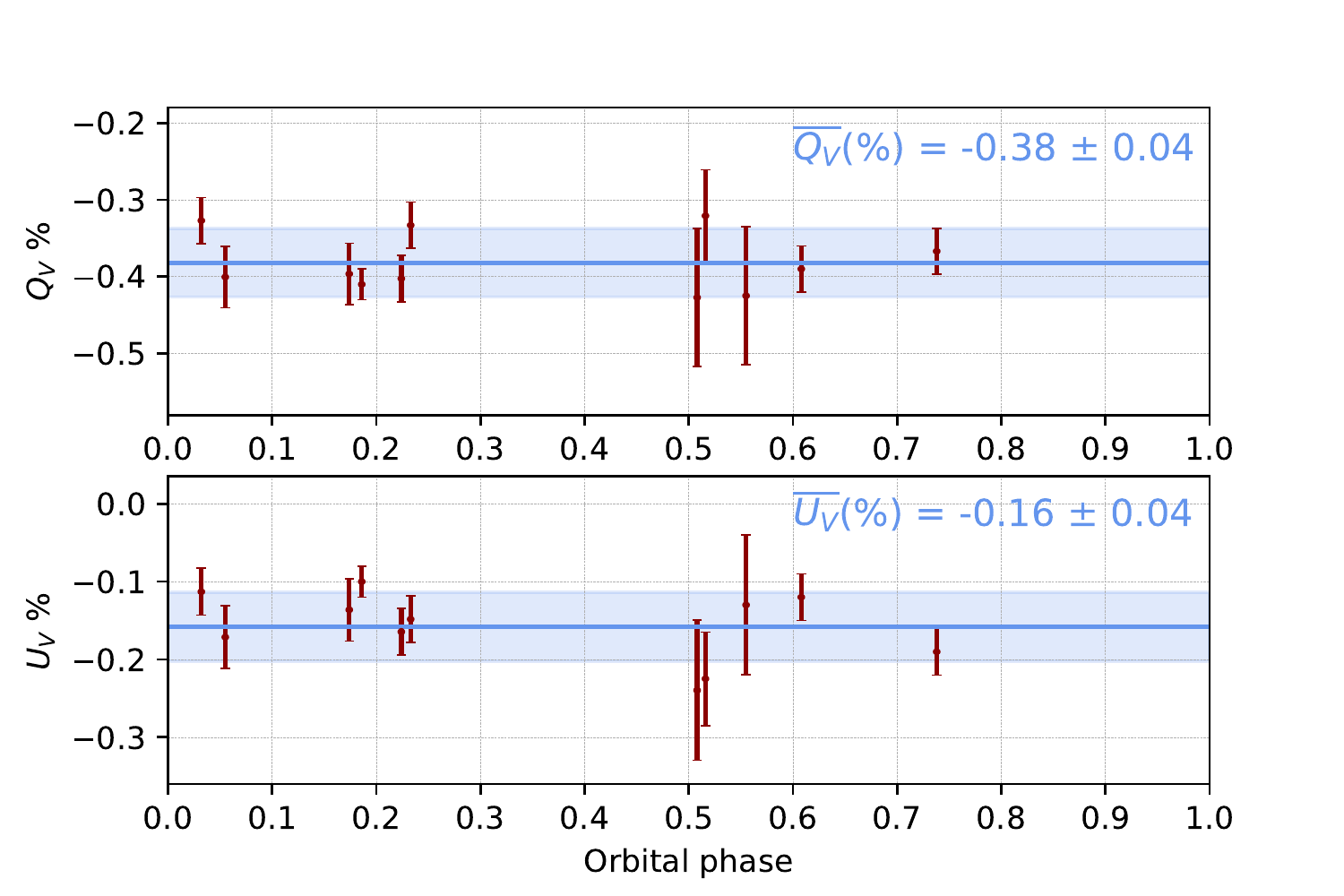}
     \end{center}
         \caption[]{Orbital phase vs. Stokes $Q_{obs}$ and Stokes $U_{obs}$ in T CrB. The circles represent the observed data in the synthetic V band. The blue line represents the average value of Stokes $Q_V$ and Stokes $U_V$ parameter respectively. The semi-translucent bands determine areas with $\pm 1\sigma$ around the average values.}
\label{orb.phase.tcrb}	 
\end{figure}

The observed value of the Stokes parameters $Q_{V}(\%)$ and $U_{V}(\%)$ in synthetic V filter obtained in different orbital phases are presented in Fig.\ref{orb.phase.tcrb}. The orbital phase is calculated using $P_{orb}= 227.5678 \pm 0.0099$ days and $T_{\circ}=2447918.62\pm0.27$, where $T_{\circ}$ is the time of maximum velocity \citep{2000AJ....119.1375F}. Solid blue lines represent the average values of the Stokes parameters $Q_V$ and $U_V$ respectively. The semi-translucent bands represent errors of the Stokes parameters. During the observations between 2018 February 17 and 2021 August 08 the variable degree of polarization is not observed.

\subsection{Interstellar polarization toward the Recurrent Nova T CrB}

The observed polarization is a vectorial sum of intrinsic polarization and interstellar polarization. When the intrinsic polarization of the observed objects is zero, the observed polarization represents interstellar polarization.  
In  the optical region, the degree of interstellar polarization
is a function of the wavelength \citep{1975ApJ...196..261S}: 
\begin{equation}
P_{ISP}(\lambda ) = P_{max}\exp (-K\ln ^{2}\frac{\lambda _{max}}{\lambda }),
\end{equation}
where $P_{max}$ is the peak 
of the interstellar polarization at wavelength $\lambda_{max}$.

In Table \ref{tab.Serkowski} is presented the parameters of the Serkowski's law for T CrB, HD 143256, BD +26 2759; HD 143161; HD 142762; HD 142053 and HD 138749. In this table are included only data with good S/N. The last two columns represent $R_V$ and grain size a. 

 \begin{table*}
\centering
\caption{The parameters of the Serkowski's law, $R_{V}$ and grain size $a$.}
\begin{tabular}{ l  c  c  c | c r   }
\hline\hline
Object       & ~~~ $P(\lambda)_{max}(\%)$  ~~  &K   & ~$\lambda_{max}($\AA$)$  & $R_{V}$$^{a}$ & $a$($\mu$m)$^{b}$   \\
\hline

\bf T CrB          & $0.46 \pm 0.01$ & $1.36 \pm 0.09$ &  ~$5194\AA \pm  89\AA$   & $ 2.9 \pm 0.2$ & 0.14  \\
\bf HD 143256      & $0.41 \pm 0.01$ & $1.15 \pm 0.11 $ &  ~$6121\AA \pm  48\AA$  & $ 3.4 \pm 0.5$ & 0.16 \\
\bf BD +26 2759    & $0.39 \pm 0.01$ & $1.28 \pm 0.22 $ &  ~$6012\AA \pm 125\AA$  & $ 3.4 \pm 0.9$ & 0.16  \\
\bf HD 143161      & $0.44 \pm 0.01$ & $1.12 \pm 0.05 $ &  ~$5686\AA \pm 44\AA$   & $ 3.2 \pm 0.4$ & 0.15  \\
\bf HD 142762      & $0.36 \pm 0.01$ & $2.01 \pm 0.02 $ &  ~$5419\AA \pm 12\AA$   & $ 3.0 \pm 0.2$ & 0.14  \\
\bf HD 142053      & $0.49 \pm 0.01$ & $1.4  \pm 0.02 $ &  ~$5324\AA \pm 13\AA$   & $ 3.0 \pm 0.2$ & 0.14  \\
\bf HD 138749      & $0.26 \pm 0.01$ & $0.76  \pm 0.02$ &  ~$5497\AA \pm 20\AA$   & $ 3.1 \pm 0.3$ & 0.15  \\

\hline
\label{tab.Serkowski}
\end{tabular} 
\\
Note: $^{a}$ - $R_{V}$ calculated with eq.\ref{eq.Rv};
 $^{b}$ - grain size calculated with eq. \ref{eq.griansize}.
\end{table*}

Interstellar polarization toward T CrB was estimated using the polarization of stars in the direction of T CrB. 
In Table \ref{tab.obsjournal} are given objects, angular distance from T CrB, distance \citep{2021AJ....161..147B}, and the last two columns represent observed degree of polarization and position angle in synthetic V filter.

\begin{figure}[htb]
    \begin{center}
      \includegraphics[width=0.4\textwidth, angle=0]{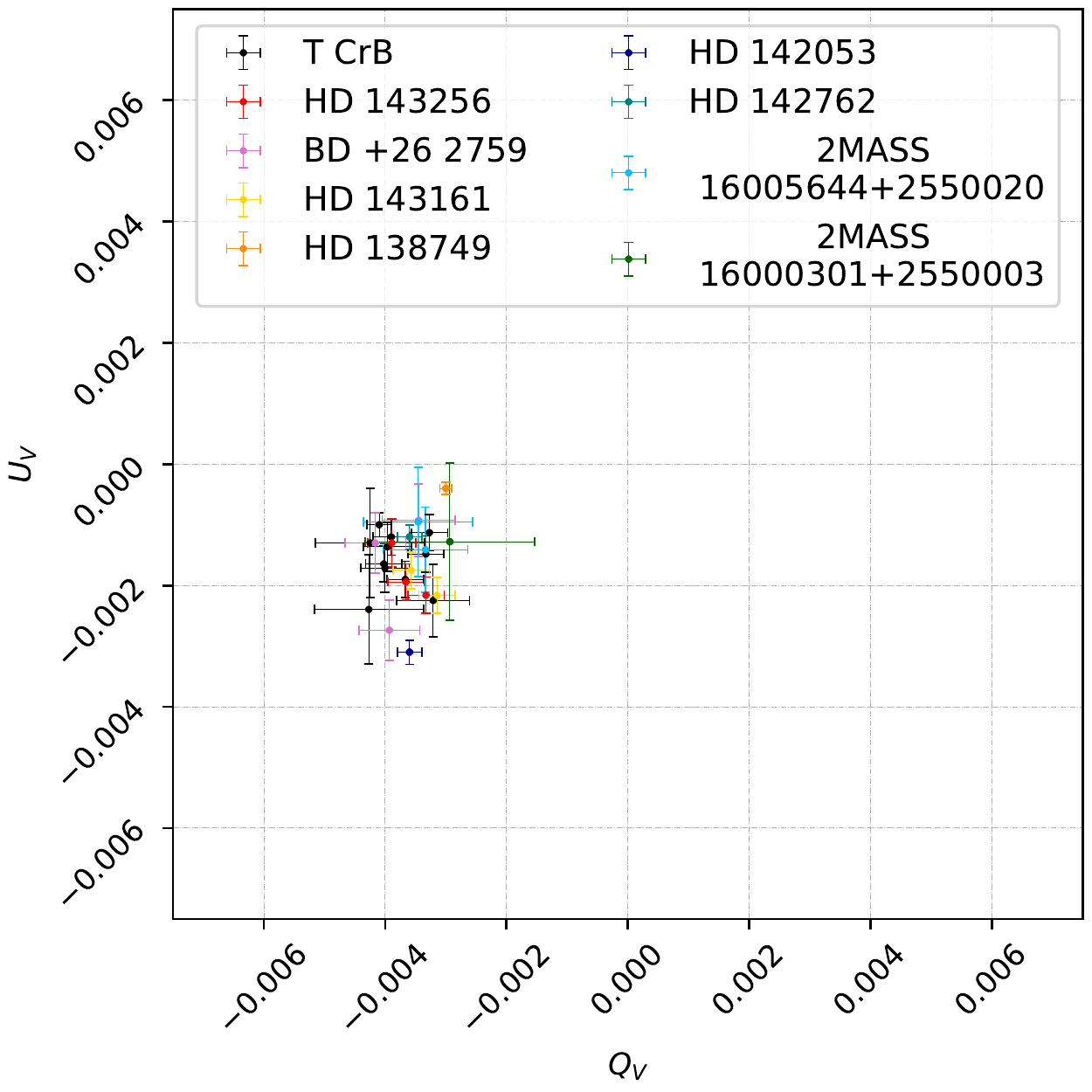}
     \end{center}
         \caption[]{QU diagram of T CrB and stars from the field of T CrB.}
	 
\label{fig.qu}
\end{figure}

\begin{figure}[htb]
    \begin{center}
      \includegraphics[width=0.5\textwidth, angle=0]{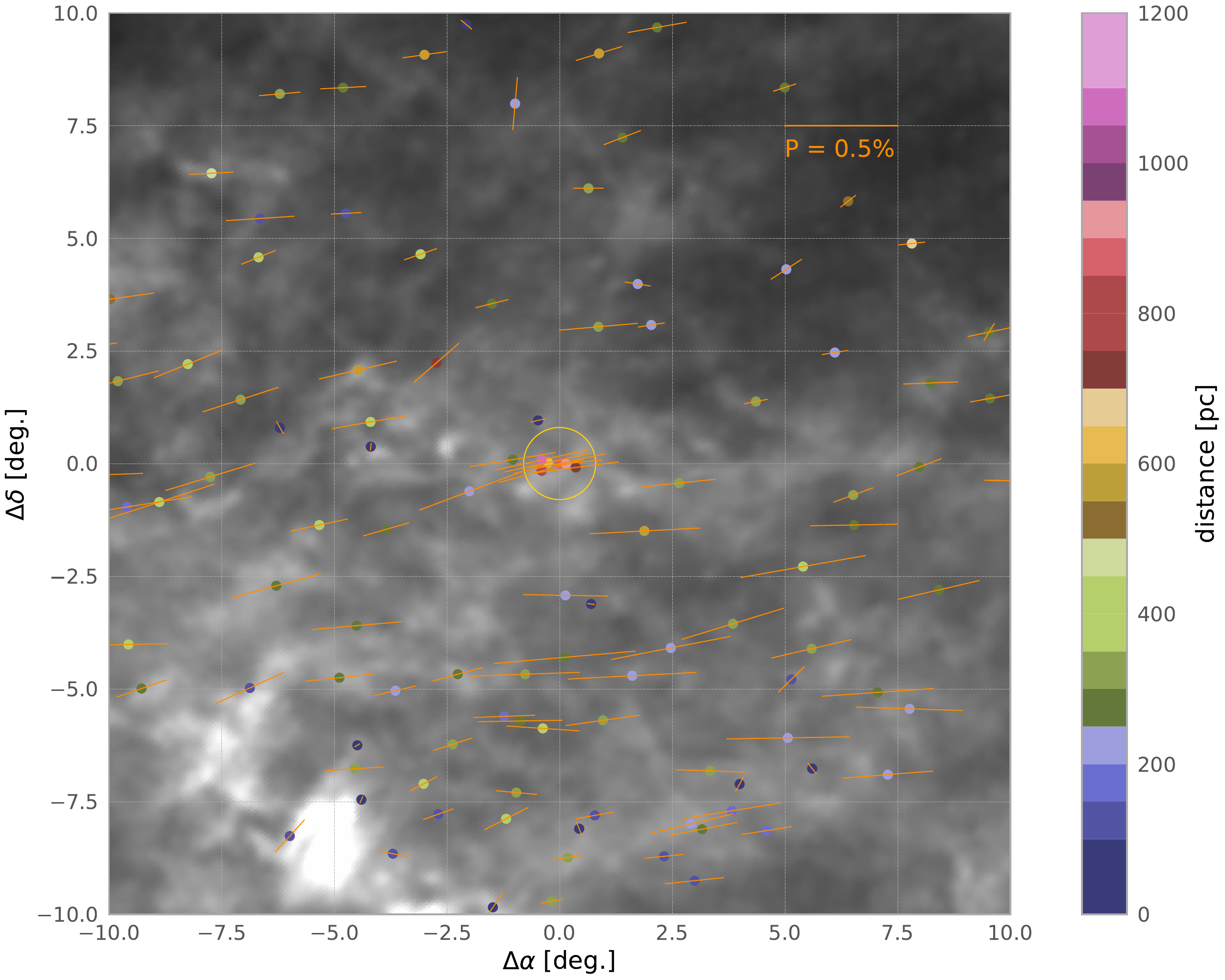}
     \end{center}
         \caption[]{The interstellar polarization of the field stars around T CrB (\citet{2000AJ....119..923H}; \citet{2014A&A...561A..24B}). The degree of polarization is proportional to the length of its bar. The horizontal bar of the top right presents 0.5\% polarization. The P.A. of the stars of the direction of T CrB (inside of the yellow circle, Table \ref{tab.obsjournal}) are similar to that of T CrB. The color of every star corresponds to its distance. The background image represents 100 $\mu$m dust emission maps \citep{1998ApJ...500..525S}.}
	 
\label{fig.2Dpol}
\end{figure}

In Fig.\ref{fig.qu} is presented a QU diagram of T CrB and stars from the field of T CrB. Observed Stokes parameters in synthetic V filter $Q_V$ and $U_V$ of T CrB have the same values as Stokes parameters $Q_V$ and $U_V$ of stars from the field of T CrB. In Fig.\ref{fig.2Dpol} are presented the observed degree of polarization and position angle of T CrB and stars from the field of T CrB obtained on 2020 February 2.
Fig.\ref{fig.qu} and Fig.\ref{fig.2Dpol} indicate that observed polarization of T CrB represents interstellar polarization.

The relationship between the grain size and the wavelength of maximum polarization for long dielectric cylinders with radius $a$ and refractive index n has been quantified by \citet{2003dge..conf.....W} as:

\begin{equation}
	\lambda_{max} \approx 2\pi a(n-1), 
\label{eq.griansize} 
\end{equation}

where $\lambda_{max}$ and $a$ are expressed in $\mu$m.
For silicate grains (n=1.6), $\lambda_{max}=5194\AA$~ yields a grain size of the order of 0.14 $\mu$m (last column of Table 3).

The linear relationship between total-to-selective extinction ratio $R_{V}$ and the wavelength of maximum polarization was suggested by \citet{1978A&A....66...57W} as: 

\begin{equation}
	R_{V}=(5.6 \pm 0.3)\lambda_{max},
\label{eq.Rv}
\end{equation}

where $\lambda_{max}$ is expressed in $\mu$m.
It is worth noting that this relationship is still debated (see \citet{2017A&A...608A.146B}). Calculated values for $R_{V}$ are presented in Table \ref{tab.Serkowski}.

\begin{figure}[htb]
    \begin{center}
      \includegraphics[width=0.5\textwidth, angle=0]{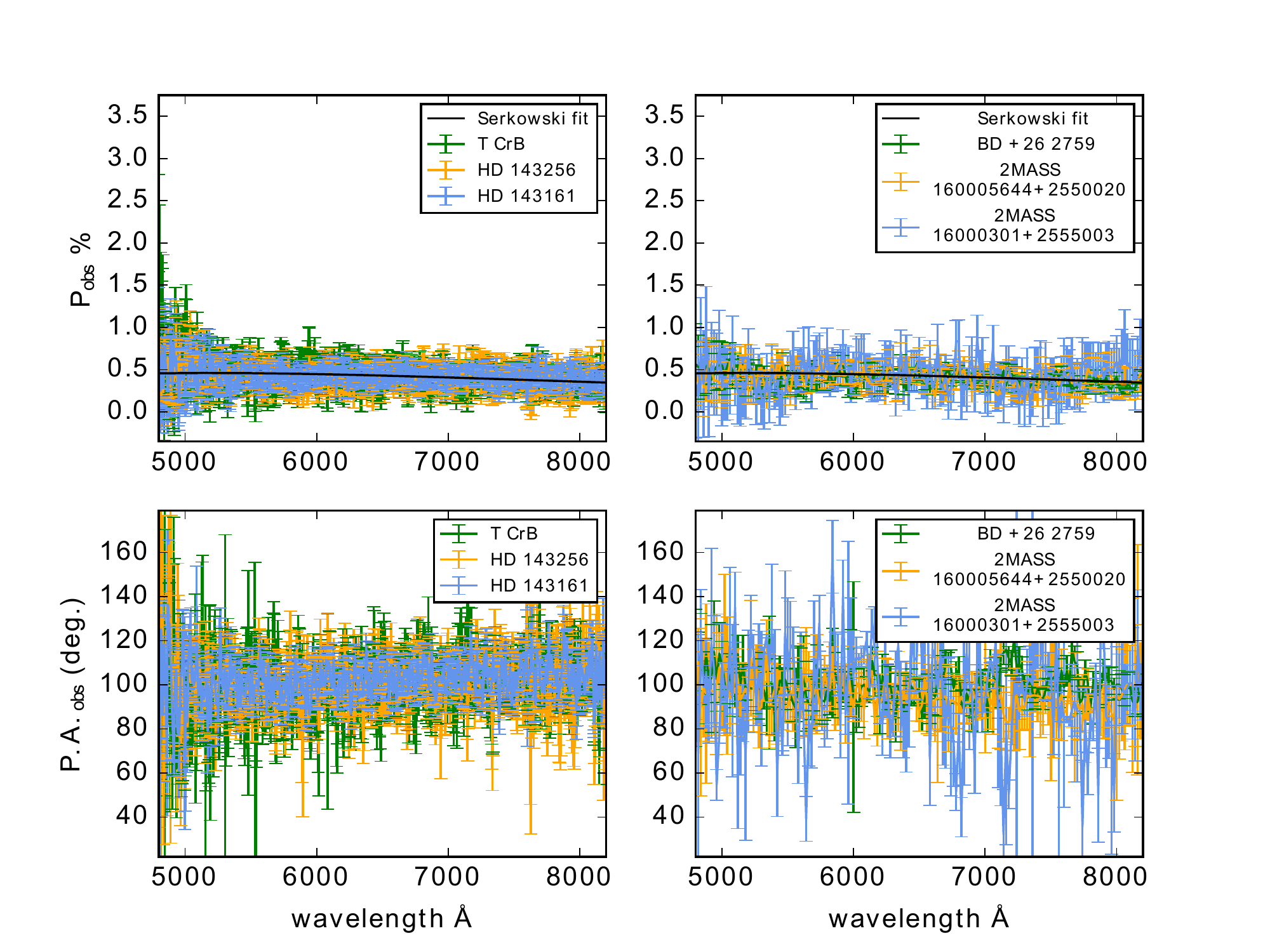}
     \end{center}
         \caption[]{Observed degree of polarization and position angle of T CrB and stars of the direction of T CrB on 2020 February 2. Black curve represents fit with Serkowski law with coefficients $P_{max}$=0.46, K=1.36 and $\lambda_{max}$=5194\AA.}

\label{fig.tcrb.polspec}
\end{figure}

\subsection{Polarization vs. distance}

The behavior of P(\%) with distance (in Kpc) is described with the equation \citep{2002ApJ...564..762F}:
\begin{equation}
P(\%) \approx 0.13+1.81d-0.47d^2+0.036d^3, 
\label{eq.fosalba} 
\end{equation}
where d is in Kpc.

\begin{figure}[htb]
    \begin{center}
      \includegraphics[width=0.4\textwidth, angle=0]{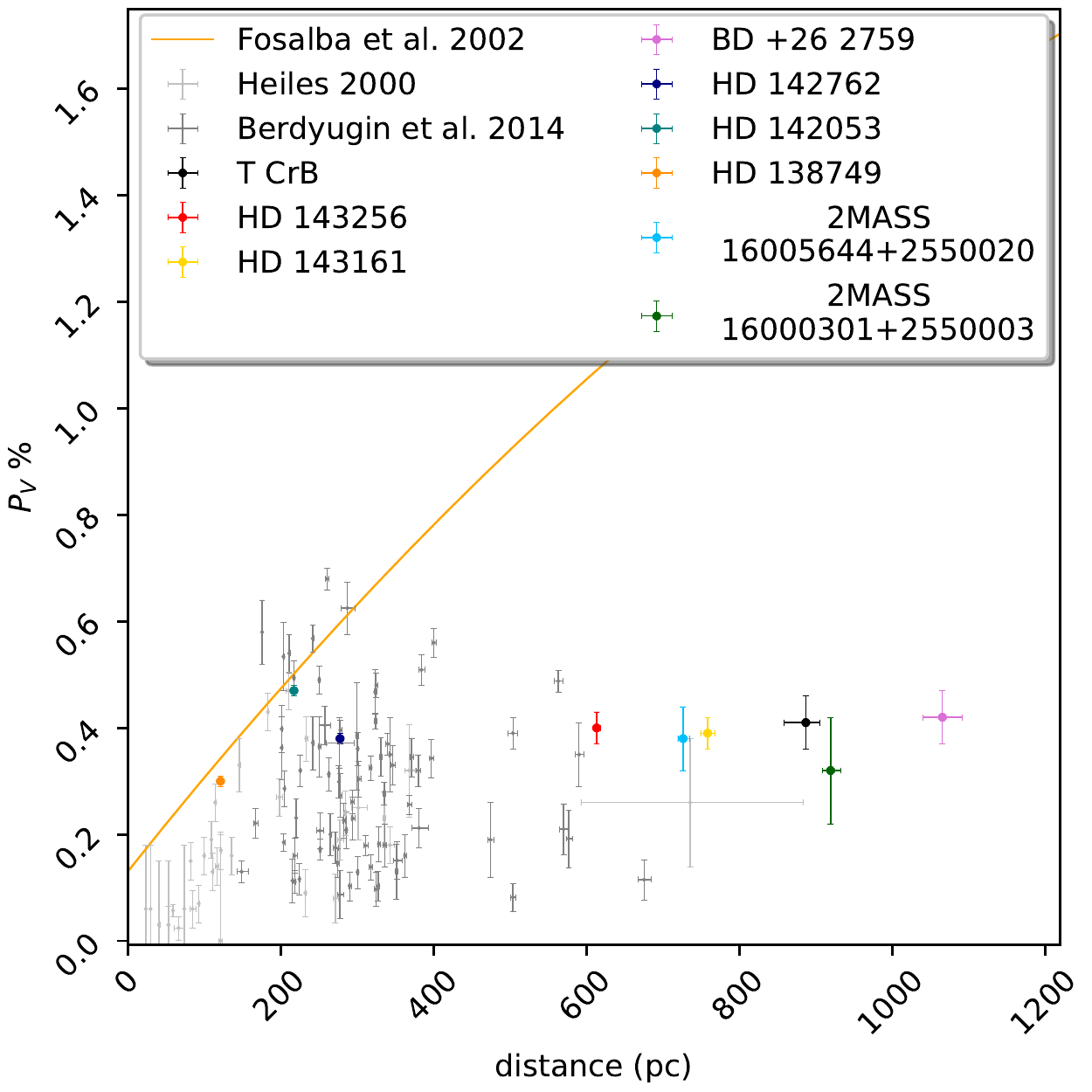}
     \end{center}
         \caption[]{Polarization vs. distance. The orange line represents the polarization vs. distance relationship \citep{2002ApJ...564..762F}.}
	 
\label{distance.cathalog}
\end{figure}

In Fig.\ref{distance.cathalog} the grey and black dots represent stars at an angular distance less than 10 degrees from T CrB. The data are taken from \citet{2000AJ....119..923H} and \citet{2014A&A...561A..24B}. Both catalogs: "Stellar polarization catalogs agglomeration" \citep{2000AJ....119..923H} and "Polarization at high galactic latitude" \citep{2014A&A...561A..24B} include relatively bright stars at the distance up to 600 pc in the field 10x10 deg. around T CrB. In Fig.\ref{distance.cathalog}  the distance to those stars is based on Gaia DR3 \citep{2021AJ....161..147B}. The orange line represents the polarization vs. distance relationship \citep{2002ApJ...564..762F} and is described with eq. \ref{eq.fosalba}. The behavior of P(\%) with the distance well describes the observed data at the distance up to 300 pc as well visible in Fig.\ref{distance.cathalog}. The scatter of the data depends on the local characteristics of the interstellar medium. This is well visible in Fig.\ref{fig.2Dpol}, where the background image represents 100 $\mu$m dust emission map \citep{1998ApJ...500..525S}.\\
The degree of polarization in the direction toward T CrB is practically constant at the distance from 400 pc to 1100 pc. The angular distance of HD 143256; BD +26 2759; HD 143161; 2MASS 16005644 +2550020 and 2MASS 16000301 +2555003 from T CrB  is in range from 7' to 24' (inside the yellow circle in Fig.\ref{fig.2Dpol}), accordingly those stars are practically in the direction of T CrB. This behavior of the degree of polarization between 400 and 1100 pc can be explained with a low-density cavity of the interstellar dust around T CrB. It can be concluded that the polarization toward T CrB is due to the foreground interstellar dust located at the distance up to $\approx$ 400 pc. 

\subsection{Comparison with catalogs}

\begin{figure}[htb]
    \begin{center}
      \includegraphics[width=0.4\textwidth, angle=0]{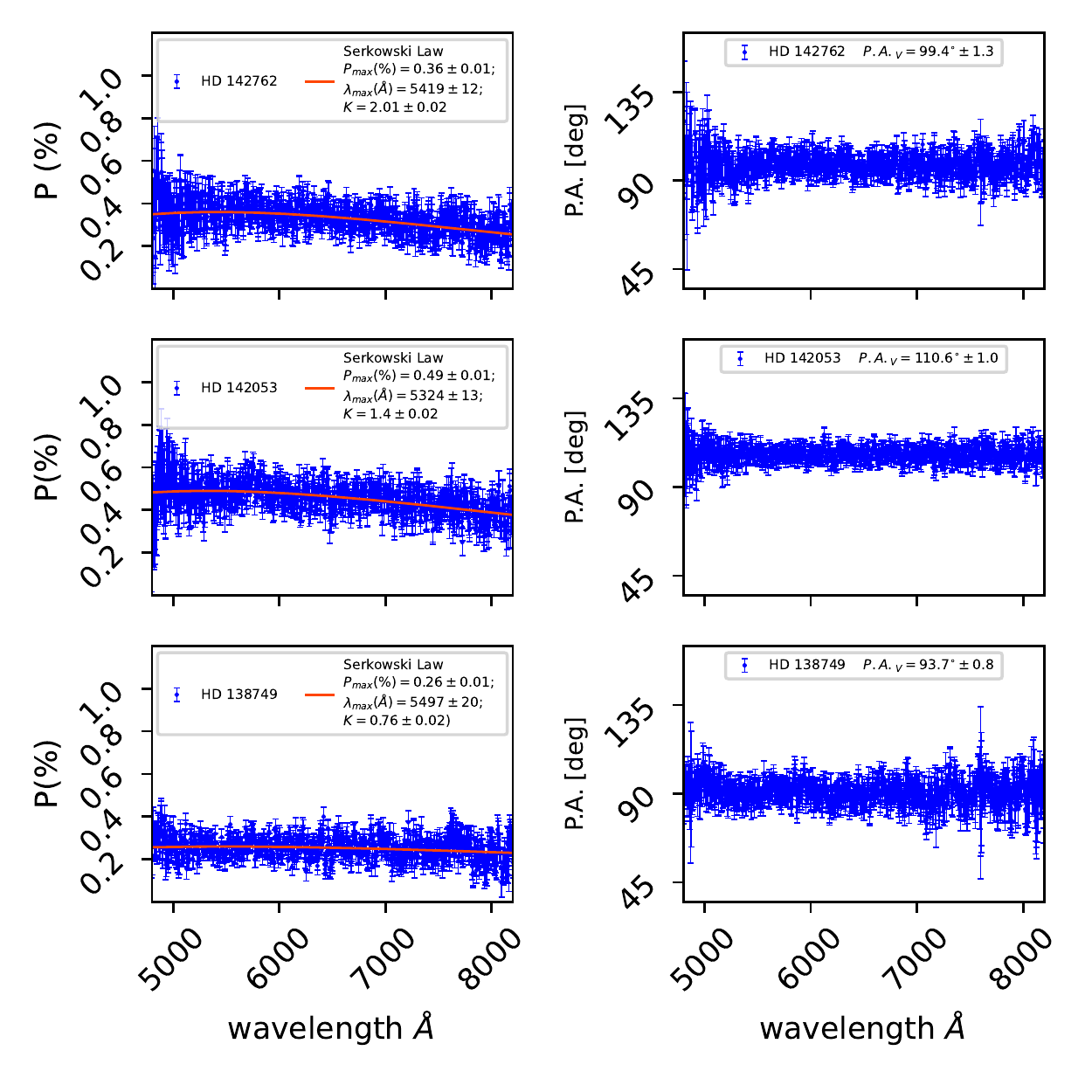}
     \end{center}
         \caption[]{Observed degree of polarization and position angle of three stars of the Heiles's and Berdyugin's catalog. The dark orange lines represent fit with Serkowski's law.}
	 
\label{fig.catalog.comparasion}
\end{figure}

For the three stars: HD 138749 \citep{2000AJ....119..923H}, HD 142053, and HD142762 \citep{2014A&A...561A..24B} were obtained spectropolarimetric observations on 2021 August 08. The synthetic V filter was used for comparison with both catalogs. The results are demonstrated in table \ref{tab.catalogs}. HD 138749 is classified as Be stars. Be stars are characterized with an intrinsic degree of polarization \citep{2010stpo.book.....C}. In the observations obtained on 2021-08-08, the $H_{\alpha}$ line is in absorption - there is no circumstellar disc around Be stars, for this reason, most probably the observed polarization represents interstellar polarization.  
In Heiles's catalog the degree of polarization of HD 138749 is $0\pm0.2\%$ and if the star has intrinsic polarization in this moment of observations the intrinsic component of polarization would be $P_{int}$ $\approx$ 0.30\% at position angle $\approx 4^{\circ}$.\\
HD 142053 and HD142762 \citep{2014A&A...561A..24B} were chosen because their P.A. is different from the general trend of the P.A. of the stars of the field of T CrB. The values of P.A. presented in this paper are very close to the general trend - in Fig.\ref{fig.2Dpol} both stars are closest to the yellow circle.
In Fig.\ref{fig.catalog.comparasion} is presented observed degree of polarization and position angle for these three stars: HD 138749; HD 142053 and HD142762. The dark orange lines represent fit with Serkowski's law. It is interesting to mention that the K coefficient of the Serkowski law is slightly different from 1.15 (the K coefficient of the original work of \citet{1975ApJ...196..261S} for two of these stars: HD 142762 and HD138749 (see Table \ref{tab.catalogs}).

 \begin{table*}
\centering
\caption{Comparison with catalogs }
\begin{tabular}{ l  c  c  c |c c| c c c  }
\hline\hline
              & Angular distance & Spectral  & $Distance^{a}$ & \multicolumn{2}{c|}{Catalogue}     &  \multicolumn{2}{c}{This work} \\               
Object        & from T CrB       & class     & [pc]           & $P_{V}$  & P.A.   &   $P_{V}$  & P.A.   \\
              & [deg.]           &           & 	              &  (\%)    & [deg.] &    (\%)	& [deg.] \\
\hline
\bf HD 142762 & $0.94$  & K0	  &$278^{+2}_{-2}$ & $0.273\pm0.041^{b}$  & $169.0\pm4.0^{b}$	 & $0.38\pm0.01$  & $99.4\pm1.3$ \\
\bf HD 142053 & $1.91$  & KIII    &$217^{+1}_{-1}$ & $0.494\pm 0.033^{b}$ & $165.0\pm2.0^{b}$	 & $0.47\pm0.01$  &$110.6\pm1.0$ \\
\bf HD 138749 & $7.96$  & B6Vnne  &$121^{+2}_{-3}$ & $0.0\pm 0.2^{c}$	  & $0.0\pm90.0^{c}$	 & $0.30\pm0.01$  & $93.7\pm0.8$  \\
\hline

\label{tab.catalogs}
\end{tabular} 
\\
Note: $^{a}$ Distance based on Gaia DR3 \citep{2021AJ....161..147B};
      $^{b}$ Polarization at high galactic latitude \citep{2014A&A...561A..24B}; 
      $^{c}$ Stellar polarization catalogs agglomeration \citep{2000AJ....119..923H}.
      The spectral class are taken from SIMBAD database.
\end{table*}

\subsection{Interstellar extinction}

Two main components of the interstellar medium are gas and interstellar dust grains. 

\subsubsection{Estimates of $E_{B-V}$ from diffuse interstellar bands (DIBs)}

The Diffuse Interstellar Bands (DIBs) represent hundreds of weak absorption features in the
wavelength range between $\sim $ 4000 and 10 000 \AA. DIBs are first mentioned in
the work of Heger \citep{1919PASP...31..304H}. Until now over 400 DiBs are registered \citep{2009ApJ...705...32H}.
The correlation between equivalent width (EW) of the DIBs and $E_{B-V}$ was suggested by \citet{2013A&A...555A..25P}.
The EW of two DIBs centered at 5780\AA~ and 5797\AA~ were used to determine $E_{B-V}$. 
In the 10x10 deg. field around T CrB has used eleven stars of the "Probing the Local Bubble with DIBs" catalog \citep{2015ApJS..216...33F}. The following equations were used to determine interstellar extinction toward those stars \citep{2013A&A...555A..25P}: 

\begin{equation}
E_{B-V} = (0.0086\pm0.0054) + (0.0023\pm0.0001)EW_{5780} 
\label{eq.puspitarini5780}
\end{equation}

\begin{equation}
E_{B-V} = (0.0203\pm0.0172) + (0.0063\pm0.0005)EW_{5797}  
\label{eq.puspitarini5797}
\end{equation}

where EW is in m\AA. 

The values of the $E_{B-V}$ are presented in Table \ref{tab.extinction.DIBs}. The Table \ref{tab.extinction.DIBs} contains: objects, $EW_{5780}$ and $EW_{5780}$ and corresponding $E_{B-V}$ obtained with equations \ref{eq.puspitarini5780} and \ref{eq.puspitarini5797}.  The last two columns represent the degree of polarization and $E_{B-V}$ calculated with equation \ref{eq.fosalba.ebv}.

\subsubsection{Estimates $E_{B-V}$ from degree of polarization}

The relationships between polarization and distance and polarization and extinction have been quantified by \citet{2002ApJ...564..762F} with the following equations: 
\begin{equation}
P(\%) \approx 0.13+1.81d-0.47d^2+0.036d^3, 
\label{eq.fosalba.distance}
\end{equation}
where  $d$ is distance in Kpc.

\begin{equation}
P(\%)\approx 3.5E(B-V)^{0.8}. 
\label{eq.fosalba.ebv}
\end{equation}

The distance to T CrB is 0.887 Kpc \citep{2021AJ....161..147B} so that using equation \ref{eq.fosalba.distance} is obtained $P(\%) \approx 0.79$. 
Toward T CrB $E_{B-V}=0.15$ \citep{1982ESASP.176..229C} so that $P(\%) \approx 0.77 $ Observed maximum of a degree of polarization of T CrB $P(\lambda)_{max}(\%) = 0.46 \pm 0.01$ is lower that calculated using eq. \ref{eq.fosalba.distance} and eq. \ref{eq.fosalba.ebv}. This is another argument in favor of low-density cavity of the interstellar dust around T CrB.

\begin{figure}[htb]
    \begin{center}
      \includegraphics[width=0.5\textwidth, angle=0]{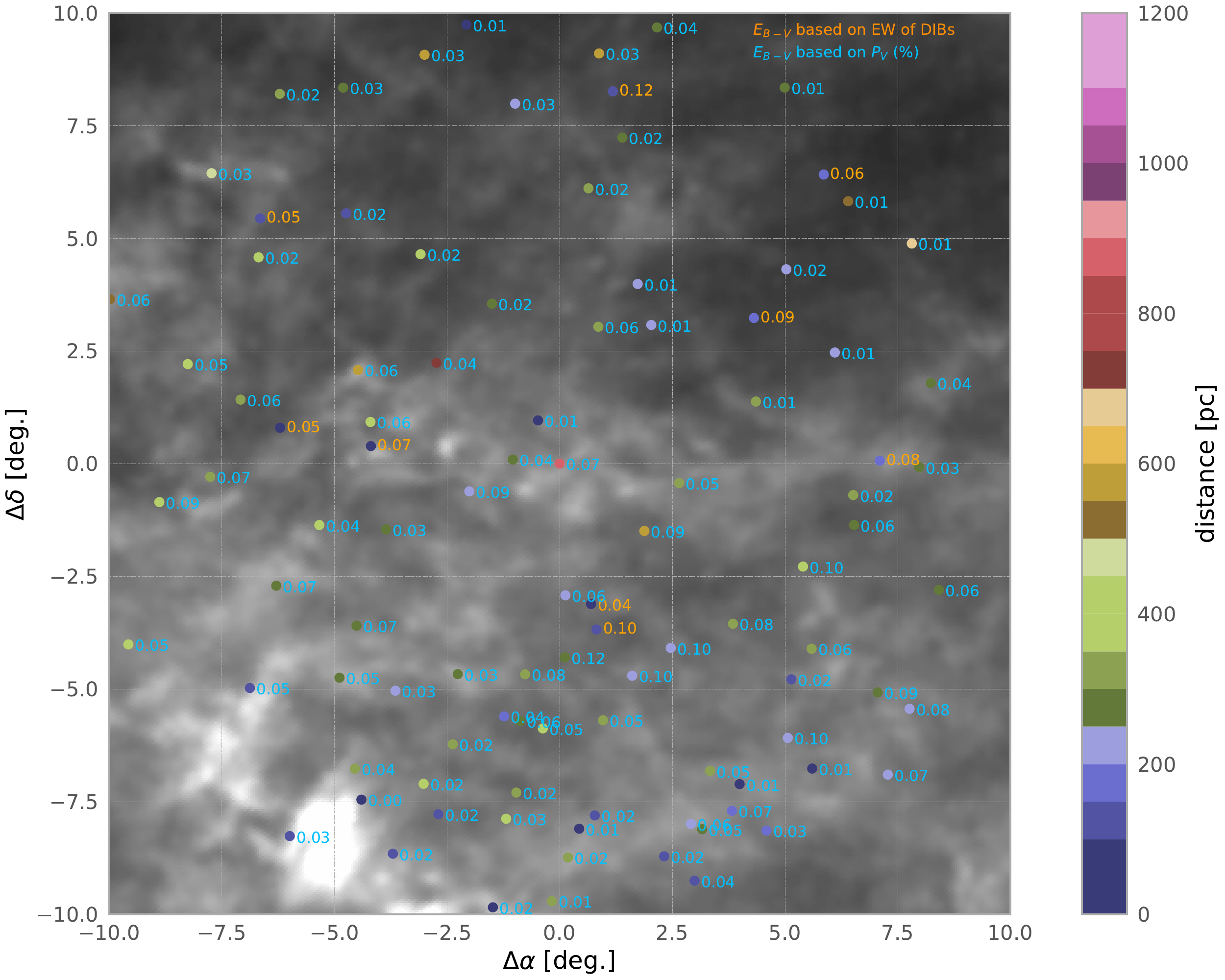}
     \end{center}
         \caption[]{The interstellar extinction of the stars in the field around T CrB. The color of every star corresponds to its distance. The background image represents 100 $\mu$m dust emission maps \citep{1998ApJ...500..525S}.}
	 
\label{fig.3D.ebv}
\end{figure}

The interstellar extinction of the field stars around T CrB is presented in Fig.\ref{fig.3D.ebv}. With blue color are presented $E_{B-V}$ obtained with equation \ref{eq.fosalba.ebv}. The degree of polarization is taken from: "Stellar polarization catalogs agglomeration" \citep{2000AJ....119..923H}, "Polarization at high galactic latitude" \citep{2014A&A...561A..24B} and this work (Table \ref{tab.obsjournal}). With orange color are presented $E_{B-V}$ based on EW of DIBs 5780\AA~ and 5797\AA. The color of every star corresponds to its distance.

 \begin{table*}
\centering
\caption{Interstellar extinction toward T CrB based on EW of DIBs and degree of polarization}
\begin{tabular}{ l  c  c  c c c c c  }
\hline\hline
Object       &   $EW^{a}_{5780}$     & $E^{b}_{B-V}$     &  $EW^{a}_{5797}$     & $E^{c}_{B-V}$  & $P_{V}$ &$E^{f}_{B-V}$	   \\	     
             &   [m\AA]             & [mag.]             &  [m\AA]	                & [mag.] & (\%)    & [mag.]  \\
\hline

\bf HD 143894	& $7.07 \pm1.71$ & $0.02\pm0.01$  & $4.89 \pm1.36$ & $0.05\pm0.03$  &                    & 		   \\
\bf HD 143992	& $41.53\pm7.42$ & $0.10\pm0.03$  & $12.42\pm3.48$ & $0.10\pm0.05$  &                    & 		   \\
\bf HD 140436	& $38.48\pm5.82$ & $0.10\pm0.02$  & $2.85 \pm1.2 $ & $0.04\pm0.03$  & $0.03\pm0.12^{d}$  & $0.003\pm0.05$   \\
\bf HD 146738	& $40.88\pm6.13$ & $0.10\pm0.02$  & $9.61 \pm2.45$ & $0.08\pm0.04$  &                    & 		    \\
\bf HD 139006	&     --         &       --       & $4.09\pm1.21 $ & $0.05\pm0.03$  & $0.06\pm0.12$      & $0.006\pm0.05$	 \\
\bf HD 148554   & $33.04\pm5.2 $ & $0.08\pm0.02$  & $7.77 \pm2.33$ & $0.07\pm0.04$  &                    & 		     \\
\bf HD 140159	&      --        &        --	  &       --       &       --       & $0.024\pm0.022^{d}$& $0.002\pm0.01$  \\
\bf HD 138749	& $25.74\pm4.37$ & $0.07\pm0.02$  & $3.03 \pm0.97$ & $0.04\pm0.02$  & $0.30\pm0.01^{e}$  & $0.046\pm0.02$    \\
\bf HD 147835	& $22.19\pm4   $ & $0.06\pm0.02$  &     --         &       --       &                    & 		    \\
\bf HD 144359	& $35.37\pm5.58$ & $0.09\pm0.02$  & $19.57\pm3.8 $ & $0.14\pm0.05$  &                    & 		     \\
\bf HD 150361   &       --       &      --        & $10.35\pm2.22$ & $0.09\pm0.04$  &                    & 		     \\

\hline
\label{tab.extinction.DIBs}
\end{tabular} 
\\
Note: $^{a}$ "Probing the Local Bubble with DIBs" \citep{2015ApJS..216...33F};  
$^{b}$ Interstellar extinction calculated with eq. \ref{eq.puspitarini5780}; 
$^{c}$ Interstellar extinction calculated with eq. \ref{eq.puspitarini5797};
$^{d}$ "Stellar polarization catalogs agglomeration" \citep{2000AJ....119..923H}; 
$^{e}$ This work (Tabe \ref{tab.obsjournal});
$^{f}$ Interstellar extinction calculated with eq. \ref{eq.fosalba.ebv};
\end{table*}

For four of the stars presented in Table \ref{tab.extinction.DIBs} we can compare the $E_{B-V}$ obtained with EW of DIBs (eq. \ref{eq.puspitarini5780} and eq.\ref{eq.puspitarini5797}) and degree of polarization (eq. \ref{eq.fosalba.ebv}). Only the star - HD 138749 has a close values of $E_{B-V}$ obtained by two methodics. 
A previous study based on spectropolarimetric observations of DIBs in the spectral range from 4480\AA~ to 6620 \AA~, \citet{2007A&A...465..899C} concluded that DIBs do not originate from grains, but they are rather large gas phase molecules that can survive in the diffuse interstellar medium. 
So the results presented in Table \ref{tab.extinction.DIBs} are not unusual and can be explained by the various causes of extinction due to the DIBs and dust.

\subsection{Spectral classification of the stars in the direction of T CrB}

\begin{figure*}[htb]
    \begin{center}
       \includegraphics[width=1.05\textwidth, angle=0]{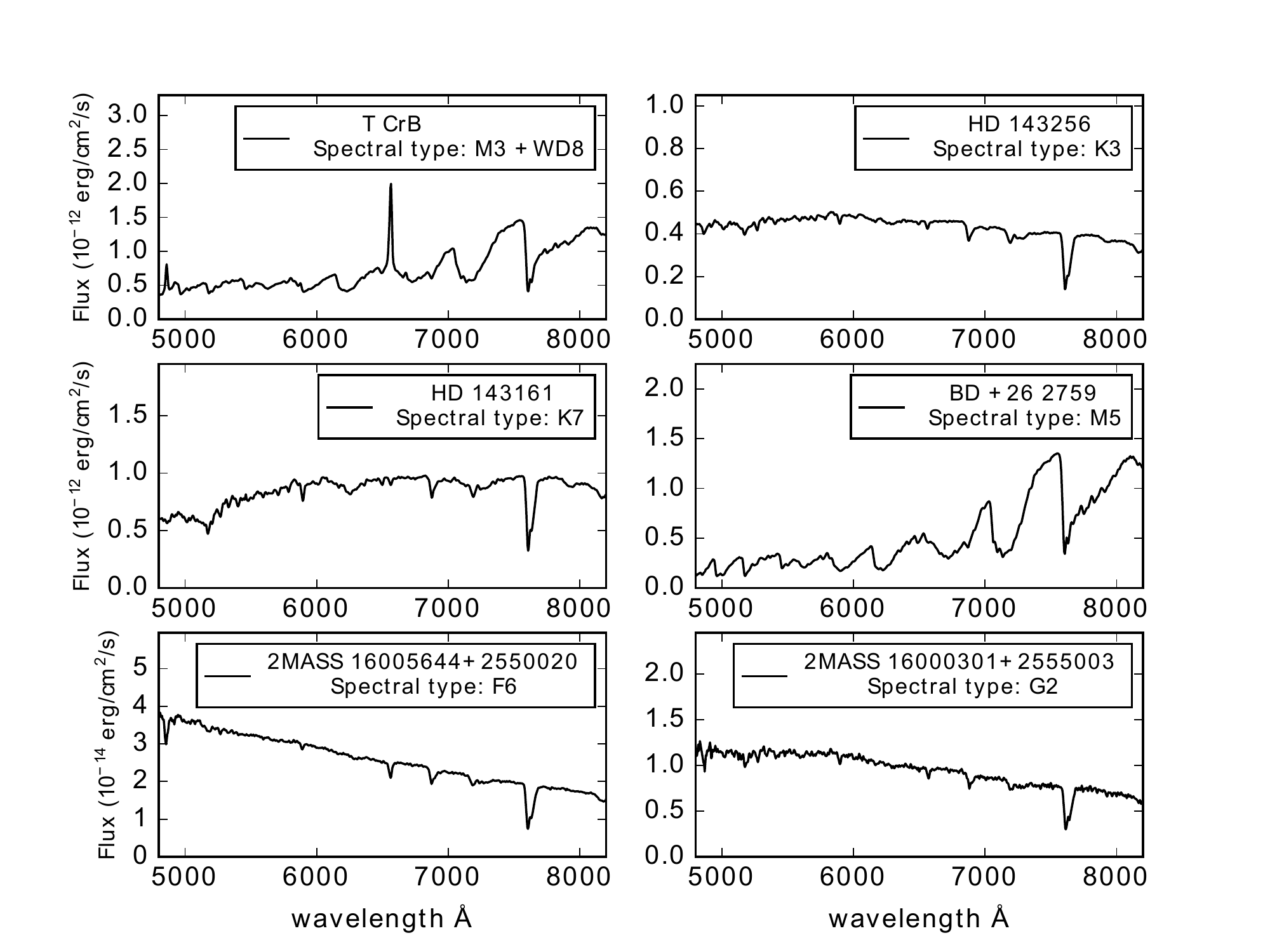}
     \end{center}
         \caption[]{Optical spectra of T CrB and the stars of the direction of T CrB were obtained in 2020-02-02. All fluxes were corrected for interstellar reddening. The absorption features around 7600 \AA~ onwards are due to the atmosphere.}
	 
\label{fig.StokesI}	 
\end{figure*}

Optical spectra (Stokes I) of T CrB and the stars of the direction of T CrB, obtained in 2020 February 2 are shown in Fig.\ref{fig.StokesI}. Spectrophotometric standard star 108 Vir was used for flux calibration. 
All fluxes were corrected for interstellar reddening of $E_{B-V}$= 0.07 using the extinction law by \citet{1989ApJ...345..245C}.
Spectral classification of observed stars has been performed with the python package PyHammer \citep{2017ApJS..230...16K}. PyHammer determines the spectral type of the object by comparing various spectral templates to the observed spectra.
The spectral type of HD 143256 is K0 and the spectral type of HD 143161 is K2 \citep{1993yCat.3135....0C}. 
For HD 143256 the best fit of spectra with PyHammer was obtained with spectral type K3. For HD 143161 the best fit of spectra with PyHammer was obtained with spectral type K7.
The spectral type of T CrB is M4 III and $M_{bol}(mag)=-2.29$ \citep{2009ApJ...697..721S}. \citet{1999A&A...344..177A} obtained a spectral type of M3-4 III for the M giant in T CrB. For T CrB the best fit of spectra obtained 2020 February 2 corresponding to spectral type M3. Using bolometric correction for this spectral type \citep{2013ApJS..208....9P} was calculated $M_{bol}(mag)=-2.02$. 

Evaluation of V-band ($m_{v}$) and $M_{bol}$ are presented in Table \ref{tab.spectralclass}. 
Table \ref{tab.spectralclass} contain: magnitude in V band ($m_{v}$), distance based on Gaia DR3 \citep{2021AJ....161..147B}, the absolute magnitude in 
V-band ($M_{V}$), Spectral type, Bolometric correction from \citet{2013ApJS..208....9P} and the last column represents absolute bolometric magnitude ($M_{bol}$). V-band magnitudes ($m_{v}$) for T CrB from the American Association of Variable Star Observers (AAVSO) database for 2020-02-02 UTC 04:43:27 and 2020-02-03 UTC 04:05:03 are $9.768\pm0.029$ and $9.962\pm0.002$, respectively. With sband IRAF procedure and using spectrophotometric standard star 108 Vir was obtained for T CrB $m_{v} = 9.9$ on date 2020-02-02.

\begin{table*}
\centering
\caption{ Spectral classification}
\begin{tabular}{ l r  c  r  c  r  r   }
\hline\hline
Object      &   $m_{v}$ &  Distance$^{a}$        &   $M_{v}$ & Spectral$^{b}$ & $BC^{c}$  & $M_{bol}$       \\
            &           &              pc      &       &        type   &        &                 \\
\hline
\bf T CrB       &    9.9    & $ 887^{+18}_{-29} $	 & 0.16     &	M3    & -1.97 &  -2.02      \\  

\bf HD 143256   &    10.1   & $ 613^{+5}_{-4} $      &  0.91      &	K3    & -0.41  & 0.50              \\

\bf BD +26 2759 &    10.6   & $ 1066^{+26}_{-26}$	 &  0.40      &	 M5   & -3.28  & -3.05       \\

\bf HD 143161   &    9.6    & $ 759^{+9}_{-9} $        & -0.10    &	K7    & -1.00  & -1.03         \\

\bf 2MASS       &    12.9   & $ 727^{+8}_{-7} $        &  3.30     &   F6    & -0.05  &  3.33    \\
\bf 16005644+2550020 &      &                 &  &       &          & \\
\bf 2MASS            &  14.1 & $ 919^{+13}_{-11}$        &  3.98    &   G2     &  -0.11  & 3.87       \\
\bf 16000301+2555003 &&&&&&\\
\hline	  							        	 				
		   
\label{tab.spectralclass}
\end{tabular} 
\\
Note:$^{a}$ - Distance based on Gaia Gaia DR3 \citep{2021AJ....161..147B}; $^{b}$ Spectral type obtained using PyHammer \citep{2017ApJS..230...16K}; $^{c}$ - Bolometric correction from \citet{2013ApJS..208....9P} Pecaut \& Mamajek (2013).

\end{table*}

\section*{Discussion}

The polarized light coming from astronomical objects brings important information for their geometry. 
Variable degree of linear polarization was observed in long period recurrent nova RS Oph \citep{1990MNRAS.243..144C} as in short period systems: U Sco (\citet{2013A&A...559A.121A}; \citet{2000A&A...355..256I}) and T Pyx \citep{2019A&A...622A.126P}.
Variable degree of polarization was detected in nova V339 Del (\citet{2017Ap.....60...19S}; \citet{2019ApJ...872..120K}) and the classical novae: V705 Cas, V4362 Sgr, V2313 Oph and BY Cir in outburst \citep{2002A&A...384..504E}. \\
In the recurrent nova RS Oph \citep{1990MNRAS.243..144C} observed variable linear polarization during 1985 outburst indicating the presence of intrinsic polarization. Observations of RS Oph at quiescence from July 2017 to July 2018 indicate that at this time, there is no intrinsic polarization in RS Oph  \citep{2019AcA....69..361N}. Similar to RS Oph at quiescence, as this study demonstrates, in T CrB also is not observed intrinsic polarization. Intrinsic degree of polarization was observed two days after the last outburst of the recurrent nova RS Oph \citep{2021ATel14863....1N}.\\
Theoretical model describes asymmetry of the ejected material in reccurent nova RS Oph \citep{2016MNRAS.457..822B}. Asymmetry in the ejected material in RS Oph after the nova eruption is well visible in near-infrared (5.5 days after 2006 outburst by \citet{2007A&A...464..119C}), optical (155 and 449 days after 2006 outburst by \citet{2007ApJ...665L..63B} and \citet{2009ApJ...703.1955R}), radio (21.5 day after 2006 outburst by \citet{2006Natur.442..279O}; 20.8 and 26.8 days after the 2006 outburst by  \citet{2008ApJ...688..559R}), 34 days after the 2006 outburst by \citet{2008ApJ...685L.137S}), and X-ray (1254 days after 2006 outburst by \citet{2021arXiv211004315M}). All observations of asymmetry indicate bipolar structure with East-West orientation. The position angle obtained with spectropolarimetry is aligned with the East-West orientation \citep{2021ATel14863....1N}.\\
Previous polarimetric observations in V filter of T CrB: $P_{V}(\%)=0.37\pm0.04$ at  $P.A._{obs}=111^{\circ}\pm3^{\circ}$ \citep{1985A&A...142..333S} are similar to reported in this paper. \citet{2016NewA...47....7M} reported that in 2015 T CrB has entered a super-active 
phase that is very similar to its state a few years before its nova outburst in 1946, however, \citet{2016MNRAS.462.2695I} argues that this is one of the numerous active phases that T CrB has experienced in the past. 
It is worth noting that dramatic change in the boundary layer in T CrB was observed by \citet{2018A&A...619A..61L}. \citet{2018A&A...619A..61L} suggest that the 'optical brightening event, which could be a similar event to that observed about 8 years before the most recent thermonuclear outburst in 1946, is due to a disk instability'. Based on an exhaustive historical optical light curve and 
the pre-eruption-plateau \citet{2019AAS...23412207S} predict that the next eruption of T CrB will be in  $2023.6 \pm 1.0$, on the other hand, \citet{2020ApJ...902L..14L} predict that T CrB is within 3-6 years of its next thermonuclear outburst. \\
Spectropolarimetric observations of RNe at quiescence and after the nova outburst allow determining interstellar and intrinsic component polarization after the nova outburst of the RNe systems. 
To obtain an intrinsic degree of polarization and position angle of the recurrent nova T CrB shorter after the upcoming outburst \citep{2020ApJ...902L..14L}, it is necessary to obtain the interstellar polarization toward T CrB. 
It can be expected that among various observing techniques spectropolarimetry first of all will give information about asymmetry after the nova outburst.

\section*{Conclusion}
The optical spectropolarimetric observation in range from 4800~\AA~ to 8200~\AA~of the recurrent nova T CrB are presented in this paper. The results indicated that:

\begin{itemize}
	\item The maximum of the degree of the linear polarization is $P_{max}(obs)(\%) = 0.46\% \pm 0.01$ at $\lambda \approx 5200$ \AA. The position angle is $P.A._{obs} = 100.^{\circ}8 \pm 0.^{\circ}9$. During the period of observations from February 2018 to August 2021 there is no intrinsic polarization in T CrB and the derived values represent the interstellar polarization. 
	\item Based on the degree of polarization the interstellar extinction toward the RNe T CrB is $E_{B-V} \approx 0.07$.
	\item The degree of polarization in the direction toward T CrB is practically constant at the distance from 400 pc to 1100 pc. This behavior of the degree of polarization between 400 and 1100 pc can be explained with a low-density cavity of the interstellar dust around T CrB. It can be concluded that the polarization toward T CrB is due to the foreground interstellar dust located at the distance up to $\approx$ 400 pc.
	\item 3D maps of the polarization and extinction of the stars in the field 10x10 deg. around T CrB are created.
	\item About the Be star HD 138749 can be expected variable degree of polarization.
\end{itemize} 
The spectropolarimetric observations at quiescence reported here can be useful to investigate intrinsic polarization after the forthcoming outburst.

\section*{Acknowledgments}

This work is supported by project number K$\Pi$-06-M$\Pi$58/1 -"Spectral and spectropolarimetric characteristics of the interstellar medium", Bulgarian National Science Fund.\\
\\
I am grateful to Antoaneta Antonova, Radoslav Zamanov, and Gerardo Juan Manuel Luna for their comments and support. I gratefully acknowledge observing grant support from the Institute of Astronomy and National Astronomical Observatory, Bulgarian Academy of Sciences.



\end{document}